\documentstyle[preprint,aps]{revtex}

\begin{document}

\title{Comment on ``Violation of the Greisen-Zatsepin-Kuzmin Cutoff:\\
A Tempest in a (Magnetic) Teapot? Why Cosmic Ray Energies above $10^{20}$
eV May Not Require New Physics''}

\draft

\author{Arnon Dar}
\address{
Department of Physics and
Space Research Institute\\
Technion, Israel Institute of Technology, Haifa 32000, Israel}
\maketitle

In a recent letter [1] with the same title, Farrar and Piran offered an
explanation for the near isotropy of the arrival directions [2] of
ultrahigh energy cosmic rays (UHECRs) and the apparent absence [3] of the
so called `GZK cutoff' in their spectrum around $10^{20}$ eV due to pion
photoproduction on the cosmic background radiation (CMB) that was
predicted independently by Greisen [4] and by Zatsepin and Kuz'min [5].
They suggested that the extragalactic magnetic fields near the Milky Way
are strong enough to deflect and isotropise the arrival directions of the
UHECRs from a few nearby sources for which their travel 
time to Earth is shorter than their attenuation time by
pion photoproduction on the CMB. They also estimated that this allows
active galactic nuclei (AGNs) or gamma ray bursts (GRBs) to be the source
of the UHECRs. However, these suggestions are inconsistent with
various observations: 

a. If our Milky Way galaxy was embedded in a $\sim \mu$G magnetic
halo/slab of the local supercluster (LSC) with dimensions larger than the
Larmor radius (eq. 5 of ref. [1]) of UHECRs, then $\sim$ GeV cosmic rays
would have been confined magnetically for times exceeding the Hubble time.
This contradicts the much smaller `age' of the bulk of cosmic
rays in the Milky Way that was inferred [6] from the measured isotopic
ratios ${\rm Al^{26}/Al^{27}}$ and ${\rm Be^{10}/Be^9}$.

b. If our Milky Way galaxy was embedded in a $\sim\mu$G magnetic halo/slab
of Mpc dimensions, diffusion in this halo/slab would have produced a quasi
uniform distribution of the bulk of the cosmic rays inside it.  This
contradicts the much smaller upper bounds on the densities of cosmic rays
at the small and large Magellanic clouds (SMC and LMC, respectively) that
were inferred from the EGRET observational limits on the diffuse gamma-ray
produced by cosmic rays interactions in the SMC and LMC [7]. 

c. Cosmic ray protons with energy smaller than the GZK cutoff are
attenuated mainly by electron-positron pair production off the CMB. Their
attenuation time is longer/comparable to the Hubble time. Thus, the
density of cosmic rays in the magnetic halo/slab must drop at the GZK
cutoff energy by a factor ${\rm \tau_h(E<E_{GZK})/\tau_{GZK}}$ where ${\rm
\tau_h(E<E_{GZK})}$ is the magnetic confinement time of UHECRs with energy
below the GZK cutoff energy and ${\rm \tau_{GZK}}$ is their
attenution time above it. Such a drop in the
intensity of UHECRs has not been observed [5]. A cosmic `fine tuning',
${\rm \tau_h(E<E_{GZK})\approx \tau_{GZK}}$, is required to `explain' why
such a drop is not observed [3].

d. If the UHECRs are isotropised by large angle magnetic scatterings, then
on the average only UHECRs that are produced within a distance smaller
than their diffusion distance during ${\rm \tau_{GZK}}$ can reach Earth.
The correct expression for their energy flux is given then by
\begin{equation}
{\rm \Phi={\Gamma(z=0)\,E_{CR}\,D\over 4\,\pi}\left({c\,\tau_h\over D_{LSC}}
\right)
\leq {\Gamma(z=0)\,E_{CR}\,D_{GZK}\over 4\,\pi}}\,
\end{equation}
where ${\rm \Gamma(z=0)}$ is the local 
rate of GRBs, ${\rm E_{CR}}$ is the GRB energy output in UHECRs,   
and ${\rm \tau_h}$ is the mean residence time of UHECRs in the magnetic 
halo/slab and the near isotropy of arrival directions of UHECRs 
requires that ${\rm D\leq D_{LSC}}$ and ${\rm \tau_h\leq\tau_{GZK}}$.
Assuming a flat Universe with ${\rm \Omega_{M}\approx 0.3}$,
${\rm \Omega_{\Lambda}\approx 0.7}$, and that the observed GRB rate 
is proportional to the star formation rate [8],
${\rm \Gamma(z=0)\approx 10^{-10}\,Mpc^{-3}\,yr^{-1}}$. 
The energy output in UHECRs of GRBs
which are powered by merger or collapse of compact stars     
is unlikely to exceed ${\rm E\approx 10^{52}\, erg}$, which is larger 
by approximately an order of magnitude than the total explosion energy  
of Type II supernova explosions. Consequently, eq. (1) yields  
${\rm \Phi\leq 3\times 10^{-3}\, eV\, cm^{-2}\,s^{-1}\, sr^{-1}}$,
which is smaller by 4 orders of magnitude than the observed flux of
UHECRs.
 
The observed evolution function of AGNs plus similar arguments exclude the
possibility that the isotropised UHECRs arrive from nearby AGNs.  However,
relativistic jets from GRBs in the Milky Way which are not pointing in our
direction can accelerate/deposit UHECRs in our Galactic halo and explain
the absence of the GZK cutoff [10]. 

\center{{\bf References}}      
\begin{enumerate}
\item 
G. R. Farrar and T. Piran, Phys. Rev. Lett. {\bf 84}, 3527 (2000). 
\item 
N. Hayashida, et al.,  Phys. Rev. Lett. {\bf 77}, 1000 (1996);
S. Yoshida, and H. Dai,  J. Phys. {\bf G24}, 905 (1998).
\item 
D. J. Bird et al., Astrophys. J. {\bf 441}, 144 (1995); 
M. Takeda,  et al. 1998, PRL, {\bf 81}, 1163 (1998).
\item 
K. Greisen,  Phys. Rev. Lett. {\bf 16}, 748 (1996).
\item 
G. T. Zatsepin and V. A. Kuz'min,  JETP Lett.  {\bf 4}, 78 (1996).
\item 
A. J. Connolly, et al.,  Astrophys. J. {\bf 486}, L11 (1997);
A. W. Strong and I. V. Moskalenko, Astrophys. J. {\bf 509}, 212 (1998)
and references therein. 
\item 
P. Sreekumar et al., Phys. Rev. Lett. {\bf 70}, 127 (1993). 
\item 
A. Dar, astro-ph 9901005, in {\it Frontier 
Objects in Astrophysics and Particle Physics}, Proc. Vulcano Workshop,
May 25-30, 1998 (eds. F. Giovannelli and G. Mannocchi) pp. 279-294.  
\item 
A. Dar, Astrophys. J, {\bf 500}, L93 (1998). 
\item 
A. Dar and A. Plaga, Astron. and Astrophys. {\bf 349}, 257 (1999).

\end{enumerate}
\end{document}